\documentclass[aps,reprint,superscriptaddress,showpacs,amssymb,prb,floatfix,preprintnumbers,longbibliography]{revtex4-1}
\usepackage{graphicx,color}
\usepackage{dcolumn}
\usepackage{amsmath}
\usepackage{amssymb}
\usepackage{epstopdf}

\usepackage[]{hyperref}
\hypersetup{colorlinks=true,linkcolor=blue,citecolor=blue,urlcolor=blue,pdfpagemode=UseNone}

\newcommand{\slrrtext}  {spin-lattice-relaxation rate}
\newcommand{\urusi}     {URu$_2$Si$_2$}
\newcommand{\urusip}    {URu$_2$Si$_{2-x}$P$_x$}
\newcommand{\slrr}      {$T_1^{-1}$}
\newcommand{\tonetinv}  {$(T_1 T)^{-1}$}

\begin{document}

\thispagestyle{myheadings}

\title{Evolution of coherence temperatures and two-fluid behavior in URu$_2$Si$_{2-x}$P$_x$, x = 0, 0.09, 0.33}

\title{NMR Investigation of antiferromagnetism and coherence in URu$_2$Si$_{2-x}$P$_x$}

\author{K. R. Shirer}
\affiliation{Department of Physics, University of California, Davis, CA 95616, USA}
\altaffiliation{Current address: Max Planck Institute for Chemical Physics of Solids, N\"{o\"{\i}}thnitzer Strasse 40, D-01187 Dresden, Germany}
\email{kent.shirer@cpfs.mpg.de}
\author{M. Lawson}
\author{T. Kissikov}
\author{B. T. Bush}
\affiliation{Department of Physics, University of California, Davis, CA 95616, USA}
\author{A. Gallagher}
\author{K.-W. Chen}
\author{R.E. Baumbach}
\affiliation{National High Magnetic Field Laboratory, Florida State University, Tallahassee, Florida 32310, USA}
\author{N. J. Curro}
\affiliation{Department of Physics, University of California, Davis, CA 95616, USA}

\date{\today}

\begin{abstract}

We report $^{31}$P and $^{29}$Si NMR in single crystals of URu$_2$Si$_{2-x}$P$_x$ for $x=0.09$ and $x=0.33$. The spectra in the $x=0.33$ sample are consistent with a homogenous commensurate antiferromagnetic phase below $T_N \sim 37$ K.  The Knight shift exhibits an anomaly at the coherence temperature, $T^*$, that is slightly enhanced with P doping.  Spin lattice relaxation rate data indicate that the density of states is suppressed for $x=0.09$ below 30 K, similar to the undoped compound, but there is no evidence of long range order at this concentration.  Our results suggest that Si substitution provides chemical pressure without inducing electronic inhomogeneity.

\end{abstract}

\pacs{76.60.-k, 75.30.Mb,  75.25.Dk, 76.60.Es}

\maketitle


The heavy fermion compound \urusi\ has captured the attention of the condensed matter physics community for more than two decades.\cite{MydoshReview} In recent years, several key experiments have shed new light on the nature of the  hidden order phase in this material.\cite{Kung2015,Tonegawa2014,BoariuPRL2013,FlouquetURS2012PRL,ChatterjeeURSPRL}  However, one of the outstanding mysteries surrounding the nature of the hidden order is how it evolves into  large-moment antiferromagnetic (AFM) order under pressure.\cite{JeffriesURSpressure,HassingerURSpressure}  Although there is a large anomaly in the specific heat and a partial gapping of the Fermi surface in the hidden order phase, there are no ordered moments.  Hydrostatic pressure or Ru substitutions (chemical pressure) gives rise to large static moments with antiferromagnetic order.\cite{Das2015,Kanchanavatee2011} Chemical pressure offers an appealing alternative to studying the properties of the system without the difficulties associated with high pressure measurements, but microscopic nuclear magnetic resonance (NMR) and muon spin rotation ($\mu$SR) measurements have revealed that the Ru substituted materials are electronically inhomogeneous, with local patches of antiferromagnetism in a hidden order background.\cite{AmatomuSRURS,KoharaURSinhomogeneity,Baek2010a} These results throw doubt on the premise that the intrinsic physics of the hidden order and AFM phases can be uncovered through systematic doping studies.

An alternative approach is to tune the ground state by Si, rather than Ru substitution.  Recently Gallagher et al. have reported the synthesis of high quality single crystals of P doped \urusi\ grown in indium flux.\cite{Gallagher2016,Gallagher2016URSPtransport} Hidden order  in \urusip\ is completely suppressed by $x=0.035$, followed by the emergence of a new phase  by $x\sim 0.3$.  Because P doping decreases the lattice size, it is natural to expect that P doping acts as chemical pressure and that the new phase is antiferromagnetic as is the case for the pure compound under pressure. It is also likely that P adds electrons, which may help to stabilize antiferromagnetism. Here we report detailed NMR studies that indicate that not only is this phase indeed AFM, but it is homogeneous and the ordered moment is comparable to that under pressure.

\begin{figure}
	\centering
	\includegraphics[width=\linewidth]{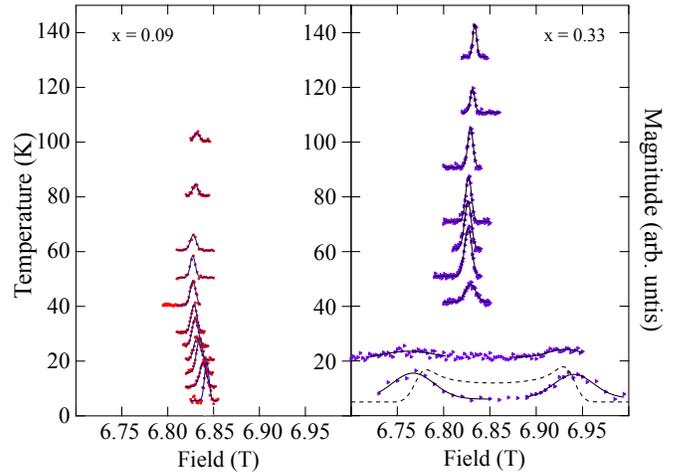}
	\caption[Waterfall plot URu$_2$Si$_{2-x}$P$_x$, x = 0.33]{\label{fig:waterfalls_doped} (color online)  Waterfall plots of URu$_2$Si$_{2-x}$P$_x$, x = 0.09, 0.33. (Left) The $x =$ 0.09 does not split, indicating there is no magnetic phase transition. (Right) The $x =$ 0.33 crystal speactra are split below $\sim$40 K and do not coexist with the central peak. The peaks show the AFM is homogeneous and commensurate. The dotted line is a model of incommensurate AFM as explained in the text. Solid lines are Gaussian fits to the spectra.}
\end{figure}

High purity single crystals of chemically substituted \urusip\ were grown using recently developed molten metal flux growth techniques described in Ref. \onlinecite{Gallagher2016}.  Single crystals of dimension $\sim 500\mu$m  and mass $\sim 0.1$ mg were selected for detailed studies with concentrations of $x=0.09$ and $x=0.33$. Magnetic susceptibility was measured using a Quantum Design SQUID magnetometer in an applied field of $H =$ 5 kOe parallel the the $c$-axis of mosaics of single crystals for temperatures $T =$ 1.8-350 K. $^{31}$P nuclear magnetic resonance (NMR) spectra ($I=1/2$, natural abundance 100\%)  were acquired by measuring spin echoes as as function of field applied parallel to the $c$-axis of single crystals at a fixed frequency of 118.315 MHz. The spin-lattice relaxation rate, \slrr, was  measured at the center of the spectra as a function of temperature. $^{29}$Si ($I=1/2$, natural abundance 4.8\%) spectra were also collected for the $x=0.09$ sample for comparison with the $^{31}$P spectrum. We estimate approximately $1\times 10^{16}$ and $4\times 10^{16}$ $^{31}$P nuclei for each sample, which exceed the number of naturally-abundant $^{29}$Si nuclei. These numbers are close to the limit of detection for NMR, thus several thousand echoes were signal averaged over several days in order to achieve sufficient signal to noise ratios.

Fig. \ref{fig:waterfalls_doped} presents a series of P NMR spectra at various temperatures at fixed frequency. In the paramagnetic phase, there is a single P resonance at frequency $f = \gamma H_0(1+K_c)$, where $\gamma =$ 17.236 MHz/T is the gyromagnetic ratio and $K_c$ is the Knight shift. The $x =$ 0.09 doping shows a smooth evolution of the spectra with no splitting of the peak  down to 4K, indicating the absence of any static internal magnetic fields.\cite{Baek2010a}
For $x =$ 0.33  the spectrum evolves smoothly down to $\sim 50$ K, below which the spectral weight is reduced due to an an enhanced spin echo decay rate, $T_2^{-1}$. This enhancement reflects the slowing down of spin fluctuations at the onset of long-range antiferromagnetic order below $T_N\sim 37$ K, in agreement with specific heat measurements.\cite{Curro2009,Gallagher2016}  The spectra at 5K and 20K at this doping reveal two separate peaks, indicating the presence of a static internal field, $\mathbf{H}_{int}$, that is either parallel or perpendicular to the applied field.  There is no evidence of a third central peak, in contrast to $^{29}$Si NMR measurements in the large moment AFM phase of \urusi\ under pressure and in URu$_{2-x}$Rh$_x$Si$_2$.\cite{Baek2010a,KoharaURSinhomogeneity}  In the latter two systems, the spectra indicated an inhomogeneous coexistence between antiferromagnetic patches and paramagnetic/hidden order regions. The absence of the central line in \urusip\ indicates a homogeneous antiferromagnetic state.

In order to investigate the nature of the distribution of internal fields, we simulate the NMR spectrum for an incommensurate spin density wave.  If the internal field is is incommensurate and varies spatially with wavelength $\lambda$ and amplitude $H_{io}$, then the theoretical lineshape is given by
\begin{equation}
P(\omega) = \frac{1}{\lambda \gamma H_{io} \sqrt{1 - \left(\frac{\omega - \gamma H_0}{\gamma H_{io}}\right)^2}}
\label{eq:theoryspectrum}
\end{equation}
where $\gamma$ is the gyromagnetic ratio and $H_0$ is the external applied field. The dotted line shown in Fig. \ref{fig:waterfalls_doped} is computed by convoluting a Gaussian with Eq. \ref{eq:theoryspectrum}. It is clear that the incommensurate model does not fit the data.  In fact, the spectral intensity vanishes between the two split peaks, and the spectrum is better fit to two Gaussians at frequencies $\gamma(H_0\pm H_{int})$.  This result implies that the antiferromagnetic order is commensurate and the internal field points along $c$.  In the AFM phase in the pure \urusi\ under pressure, neutron scattering reveals ordered U moments oriented along the $c$-direction with wavevector $\mathbf{Q}_{AF} = (1,0,0)$.\cite{Amitsuka2007}

\begin{figure}
	\centering
	\includegraphics[width=0.8\linewidth]{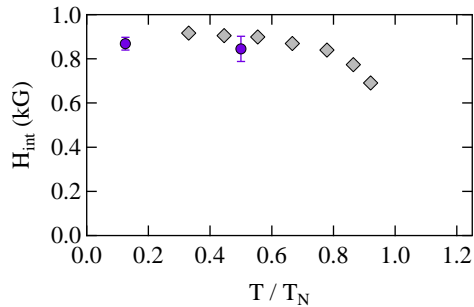}
	\caption[H$_{int}$ for URu$_2$Si$_{2-x}$P$_x$, x=0.33 and pure \urusi\ at 8.3 kbar]{\label{fig:Hint_x33_pure} (color online) The internal field at the P in URu$_2$Si$_{2-x}$P$_x$ with x = 0.33 ($\bullet$, blue) and at the Si site in the pure compound ($\blacklozenge$, gray) in the large moment antiferromagnetic phase. Data for the latter are reproduced from Ref. \onlinecite{KoharaURSinhomogeneity}.}
\end{figure}

The  internal field, $H_{int}$ = $\Delta f/\gamma$ where $\Delta f$ is the peak splitting,  is shown in Fig. \ref{fig:Hint_x33_pure}.  Similar values were reported at the Si site in \urusi\ under pressure.\cite{KoharaURSinhomogeneity}  If we assume a simple relationship between the ordered moment and the internal field: $H_{int} = B S_0$, where $B$ is the hyperfine coupling and $S_0$ is the ordered U moment, and use a hyperfine coupling of 4 kOe/$\mu_B$ (see below), we estimate an ordered moment of $0.22\pm 0.01 \mu_{B}$/U.  This value is comparable to the the ordered moment observed in the large moment AFM phase of the pure system,\cite{KoharaURSinhomogeneity} and in the Rh-doped material.\cite{Baek2010a}  This value, however, is less than that reported by neutron scattering which find $S_0\sim0.4\mu_B$/U.\cite{Amitsuka2007} The discrepancy is likely related to a more complex hyperfine coupling between the Si/P nuclei and the five nearest neighbor U moments.


\begin{figure}
	\centering
	\includegraphics[width=\linewidth]{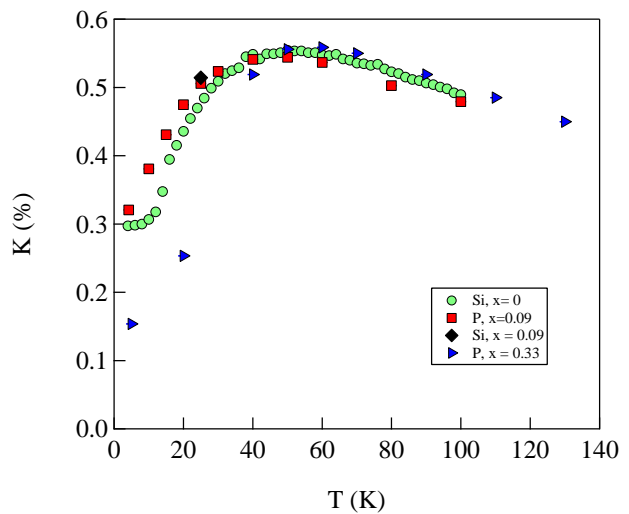}
	\caption[FWHM in \urusip]{\label{fig:KvsT} (color online) The P and Si Knight shifts in \urusip\ for $x=0$, 0.09, and 0.33. Data for the pure compound is reproduced from Ref. \onlinecite{ShirerPNAS2012}, but has been scaled by 0.8475 to agree with the results in the single crystal.\cite{ShirerURSpressure}}
\end{figure}

The spectra in Fig. \ref{fig:waterfalls_doped} show a clear evolution of the Knight shift, which is shown in Fig. \ref{fig:KvsT}.  The Knight shift in the pure compound (acquired in an aligned powder) has been scaled to match our previous results in single crystals.\cite{ShirerPNAS2012,ShirerURSpressure}  For the $x=0.09$ sample, both Si and P were measured at 25K, and the data coincide, as seen in Fig. \ref{fig:KvsT}. This fact is important because it indicates that both the Si and the P are probing the same physics, and it is appropriate to compare data measured at the two crystallographically identical sites. All three samples exhibit similar trends, with maxima around 50 K.  For the $x=0.33$ sample the shift in the antiferromagnetic state was determined by the average of the split resonances.

\begin{figure}
	\centering
	\includegraphics[width=\linewidth]{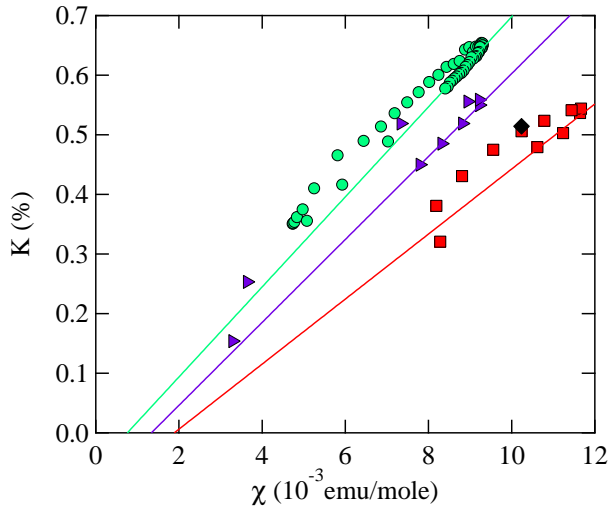}
	\caption[K vs $\chi$ for different dopings of URu$_2$Si$_{2-x}$P$_x$]{\label{fig:doped_KS_chi} (color online) Knight shift versus the magnetic susceptibility in URu$_2$Si$_{2-x}$P$_x$ with x = 0, 0.09, and 0.33; symbols are identical to Fig. \ref{fig:KvsT}.  Solid lines are fits to the high temperature data as described in the text.}
\end{figure}

Figure \ref{fig:doped_KS_chi} shows the Knight shift versus susceptibility with temperature as an implicit parameter.  In \urusi, as in most heavy fermion systems, the Knight shift arises because the nuclear spins $\hat{\mathbf{I}}$ couple both to the itinerant electron spins, $S_c$, and to the localized $f$ moments, $S_f$: $\mathcal{H}_{hyp} = A\hat{\mathbf{I}}\cdot\mathbf{S}_c + B\hat{\mathbf{I}}\cdot\mathbf{S}_f$, where $A$ and $B$ are the hyperfine couplings.\cite{ShirerPNAS2012}  The Knight shift is given by: $K = A\chi_{cc} + (A+B)\chi_{cf} + B\chi_{ff}$ and the susceptibility is: $\chi=\chi_{cc}+2\chi_{cf} + \chi_{ff}$. At high temperatures, both $K$ and $\chi$ are dominated by $\chi_{ff}$, and therefore are proportional to one another, as seen in Fig. \ref{fig:doped_KS_chi}.  The solid lines are linear fits to the high temperature data, and the fit parameters are summarized in Table \ref{tab:hypandT*}. The slope of the high temperature linear fit yields the transferred hyperfine coupling, $B$, and the intercept $K_0$ is a temperature independent offset that is usually given by the orbital susceptibility and diamagnetic contributions.\cite{NissonCEFSOC2016}  It is noteworthy that the hyperfine coupling to the P is nearly identical to the coupling to the Si, which suggests that the local electronic structure is not significantly perturbed by the presence of the dopant.  The variation in the $K_0$ parameter may also be related to errors in measurement of the susceptibility due to the small signal of the crystals which have very low masses.

 \begin{table}
 \caption{\label{tab:hypandT*} Doping, hyperfine parameters, and coherence temperatures in \urusip.}
 \begin{ruledtabular}
 \begin{tabular}{lllll}
 x     & nucleus & B (kG/$\mu_{B}$) & $K_0$ (\%) & $T^*$ (K)  \\ \hline
0     & $^{29}$Si & 4.34$\pm$0.10 & $-0.08\pm02$ & 73.1$\pm$0.5   \\
0.09  & $^{31}$P &3.05$\pm$0.07  & $-0.10\pm 0.14$  & 74.2$\pm$1.6    \\
0.33  &$^{31}$P & 3.89$\pm$0.08 & $-0.09\pm 0.01$ & 73.2$\pm$0.3  \\
 \end{tabular}
 \end{ruledtabular}
 \end{table}

\begin{figure}
	\centering
	\includegraphics[width=\linewidth]{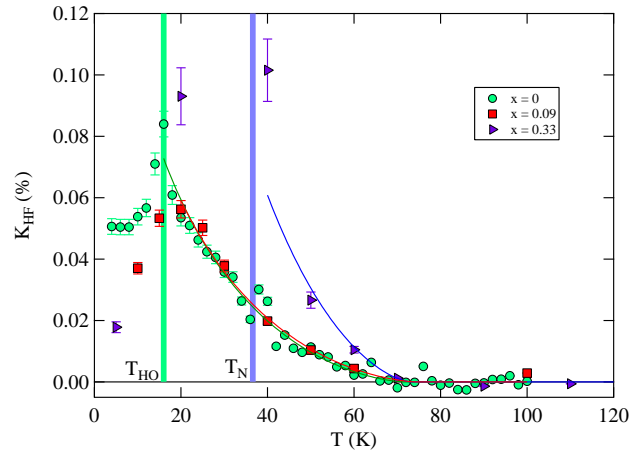}
	\caption{\label{fig:Khf_all} (color online) $K_{HF}$ versus temperature in URu$_2$Si$_{2-x}$P$_x$ with x = 0 ,0.09, and 0.33; symbols are identical to those in Fig. \ref{fig:KvsT}. Solid lines are fits to Eq. \ref{eqn:YangPines} as described in the text.  The dashed vertical lines indicate $T_{HO}$ for $x=0$, and $T_N$ for $x=0.33$.}
\end{figure}

Below the coherence temperature, $T^*$, the linear relationship between $K$ and $\chi$ breaks down as $\chi_{cf}$ grows in magnitude.\cite{ShirerPNAS2012,ROPP2016} This quantity reflects the heavy electron component in the two-fluid model that emerges due to collective hybridization.\cite{Yang2014,YangPinesPNAS2012,Yang2011,YangPinesNature,YangDavidPRL}  To examine the temperature dependence of the heavy electron susceptibility, we compute $K_{HF} = K-K_0 - B\chi = (A-B)(\chi_{cf} + \chi_{cc})$, as shown in Fig. \ref{fig:Khf_all}. It has been shown that $K_{HF}$ scales universally with $T/T^*$, where $T^*$ is material dependent, and that $T^*$ agrees well with several other experimental measurements of the coherence temperature of the Kondo lattice.\cite{Curro2004}
The solid lines in Fig. \ref{fig:Khf_all} to the empirical two-fluid expression:
\begin{equation}
K_{HF}(T) = K_{HF}^0(1-T/T^*)^{3/2}[1+\log(T^*\!/T)]
\label{eqn:YangPines}
\end{equation}
where $K_{HF}^0$ is a constant.  $T^*$ is given in Table \ref{tab:hypandT*}. We find that $T^*$ agrees well with resistivity measurements, which reveal only a modest $\sim 5$\% increase in $T^*$ over this doping range.\cite{Gallagher2016URSPtransport} Surprisingly, however, $K_{HF}$ decreases below 20K for the $x=0.09$ sample.  This `relocalization' phenomenon has been observed previously in other heavy fermion materials and may reflect a precursor to the emergence of local moment order at higher dopings.\cite{Warren2011,ShirerPNAS2012}

\begin{figure}
	\centering
	\includegraphics[width=\linewidth]{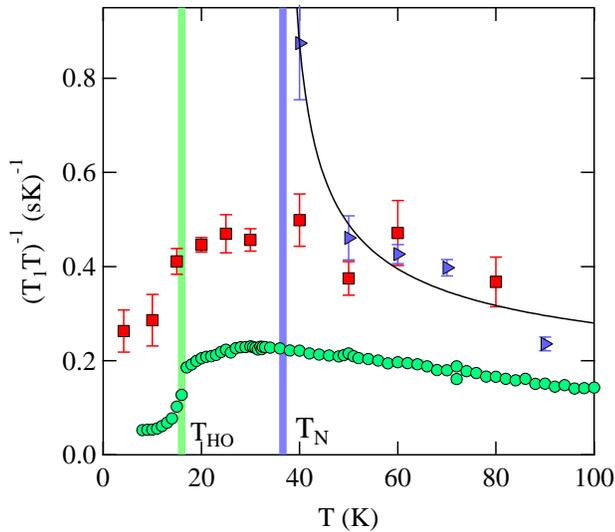}
	\caption{\label{fig:T1T_all} (color online) $(T_1T)^{-1}$ versus temperature for the Si at $x=0$, and the P at $x=0.09$ and 0.33. Data symbols are identical to Fig. \ref{fig:KvsT}.  The data for $x=0$ is reproduced from Ref. \onlinecite{ShirerPNAS2012}, and has been scaled by $(^{31}\gamma/^{29}\gamma)^2$ for comparison. The solid line is a fit to the temperature dependence in the $x=0.33$ data as described in the text.}
\end{figure}

We have also measured the \slrrtext\ as a function of temperature at the P site, as shown in Fig. \ref{fig:T1T_all}, which  includes data for the Si in pure \urusi.  Both $^{29}$Si and $^{31}$P are spin 1/2 nuclei, and the magnetization recovery data were well fit by the standard recovery function, $M(t) = M_0(1- f e^{-t/T_1})$, where $M_0$ is the equilibrium magnetization and $f$ is the inversion fraction. The spin lattice relaxation rate, \slrr, is proportional to the square of the gyromagnetic ratio, therefore we have scaled the Si data for the $x=0$ compound by the ratio $(^{31}\gamma/^{29}\gamma)^2$ for comparison.  For the $x=0.09$, $(T_1T)^{-1}$ exhibits a maximum around 30K, similar to the pure compound. Below this temperature, $(T_1T)^{-1}$ is reduced by approximately 50\% without any evidence for long range order.  This behavior suggests a partial suppression of the density of states, as observed in pure \urusi. In the latter, $(T_1T)^{-1}$ is suppressed by approximately 20\% before hidden order develops.  This result suggests that the suppression is unrelated to precursor fluctuations of the hidden order, but may be related to a pseudogap associated with the background spin fluctuations.\cite{ShirerURSPRB2012}

For the $x =$ 0.33 sample, \tonetinv\ exhibits a significant enhancement above $T_N$ due to the critical slowing down of spin fluctuations.  We find that the data fits well to the form: $(T_{1}T)^{-1} = a + b/\sqrt{T - T_{\text{N}}}$, an expression appropriate for weak itinerant antiferromagnets.\cite{Moriya1974,Curro2001}  The best fit is shown by the solid line in Fig. \ref{fig:T1T_all} with $T_\text{N}$ = 36.6$\pm$1.8 K.  The enhanced spin lattice relaxation associated with this critical slowing down also affects the spin-spin decoherence time, $T_2$, which is responsible for the low signal to noise evident in Fig. \ref{fig:waterfalls_doped} just above $T_N$.

The relaxation rate for the P in the \urusip\ is nearly twice as large as the Si in the pure \urusi, even after scaling by the square of the gyromagnetic ratios.  This difference could be related to form factors and hyperfine coupling differences to the P and the Si.\cite{DioguardiPdoped2015} It is also possible that the spin fluctuations are higher in the P doped samples, however the temperature dependence of \tonetinv\ above 50 K is nearly identical for all three systems, which suggests otherwise.  Furthermore, \tonetinv\ has been observed to decrease under pressure in the pure \urusi.  A third possibility is that the local density of states is slightly different at the P site.


In summary, we have performed $^{31}$P NMR measurements in single crystals of \urusip\ with $x = $ 0.09 and 0.33. For the AFM $x = $0.33 crystal, we determined $T_\text{N}$ = 36.6$\pm$1.8 K with a commensurate internal field $H_{int}\sim 0.85$ kOe oriented along the $c$ direction at 5 K. This behavior is entirely consistent with the large moment AFM phase in pure \urusi\ under pressure. Furthermore, we find that the AFM phase in \urusip\ is homogeneous, in contrast to the heterogeneous patches of AFM  observed in the \urusi\ under pressure and U(Ru,Rh)$_2$Si$_2$. We find the $x =$ 0.09 crystal undergoes no phase transition, but the \tonetinv\ data for this crystal reveal a partial suppression of the density of states. All three compounds exhibit a Knight shift anomaly at $T^*\sim 72$K.  For the $x=0.09$ crystal, the heavy electron component of the Knight shift is reduced below 20 K, suggesting a relocalization of the moments as the system is tuned toward long range antiferromagnetic order for sufficiently large P doping. Our results indicate that P doping offers an avenue to tune the ground state properties cleanly, without inducing an inhomogeneous electronic response.

We thank Y-f. Yang and D. Pines for enlightening discussions, and P. Klavins for assistance in the laboratory. Work at UC Davis was supported by the NSF under Grant No.\ DMR-1506961. Work performed at the National High Magnetic Field Laboratory (NHMFL) was
supported by National Science Foundation Cooperative Agreement No. DMR-0084173, the State of Florida and the DOE. A portion of this work was
supported by the  NHMFL User Collaboration Grant Program (UCGP).


\bibliography{Shirer_URS_P_doped}

\end{document}